\documentclass[twocolumn,prb,showpacs,preprintnumbers,amsmath,amssymb]{revtex4}
\usepackage{array}
\usepackage{amsbsy}
\usepackage{graphicx}% Include figure files
\usepackage{epsf}% Include figure files

\begin{document}

\title{Transport properties of nano-devices: a one dimensional model study}
\author{Zhongxi Zhang, C. F. Destefani, Chris McDonald, and Thomas Brabec}
%\email{zhongxi.zhang@science.uottawa.ca}
\affiliation{Center for Photonics Research, University of Ottawa, Ottawa,
K1N 6N5 ON, Canada}% \\
\date{\today}
%\maketitle

\begin{abstract}
A 1D model study of charge transport in nano-devices is made by comparing multi-configuration time dependent
Hartree-Fock and frozen core calculations. The influence of exchange and Coulomb correlation on the tunneling
current is determined. We identify the shape of the tunneling barrier and the resonance structure of the
nano-device as the two dominant parameters determining the electron transport. Whereas the barrier shape
determines the size of the tunneling current, the resonances determine the structure of the current.
%I-V diagram.
\end{abstract}
\maketitle

Nano-systems have been receiving great attention lately, mainly because of two reasons. First, they allow the
realization of nano-scale electronic devices, such as Coulomb blockade structures, transistors, diodes, and
switching devices. \cite{Porath,Park,Ventra00,Wong,Chen,Ventra01,Tian98} Second, transport properties are
governed by quantum mechanics, which opens a novel venue for investigating fundamental transport properties
of quantum few-/many-body systems. The theoretical description of quantum transport properties has proven to
be a true challenge. Conventional theories, such as the Landauer-Buttiker
approach and density functional
theory do not perform well, %\cite{Lang00,Taylor01,Damle01,Aviram,Reed,Toher}
\cite{Lang00,Taylor01,Damle01,Reed,Toher} 
and are orders of magnitude off the experimentally measured currents.
Recently, configuration-interaction (CI) analysis was shown to give a much
better agreement, \cite{Delaney04} with currents about three times the experimental values.

Despite this progress, the physics underlying electron transport in nano-devices is not yet well understood.
Here, a first step is made to close this gap. Our analysis is based on a one dimensional multi-configuration
time-dependent Hartree-Fock (MCTDHF) method \cite{Zanghellini04}. The MCTDHF approach takes full account of
exchange and correlation effects. Whereas in CI the basis is fixed and only the expansion coefficients are
optimized, in MCTDHF both coefficients and basis functions are optimized. As a result, a much smaller basis
set can be used. This proves important especially for transport processes in continuum-like structures, where
a CI analysis requires a large number of basis functions for convergence.

The influence of many-body effects on transport phenomena presents a central issue in the characterization of
nano-devices. We use the following approach to investigate this question. The MCTDHF analysis is compared to
time-dependent Hartree-Fock (TDHF), frozen-core Hartree-Fock (FCHF) and frozen core Hartree (FCH) calculations.
Comparison of the different approaches reveals the role of exchange and correlation in the electron transport.
It also gives a sense of the quality of various approximations used in the electron transport analysis of
nano-devices. Further, a correlation measure is introduced with the goal to quantify the influence of many-body
effects. Finally, we identify two parameters that determine the tunneling characteristics, namely, the shape of
the barrier and the resonance structure of the nano-device.

\begin{figure}
  \begin{center}
     \includegraphics[width=8.5cm,height=6.0cm]{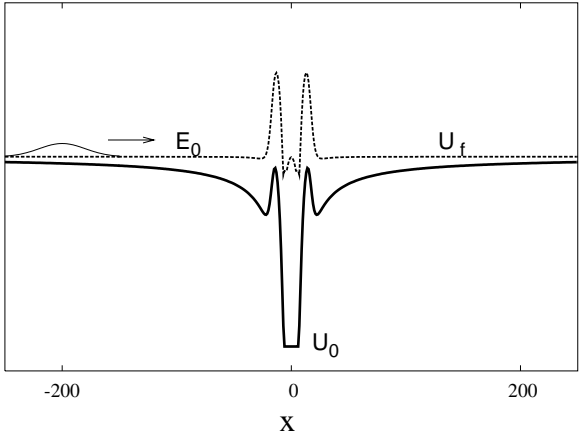}%{2cw2eldsb.eps}
     \caption{
              The effective ion potential $U(x)$ (thick solid),
              the initial Gaussian wavepackect (thin solid),
              and the effective frozen Hartree potential
              $u_{eff}^{FCH}(x)=U(x)+u^{FCH}(x)$ (dashed).
              The Coulomb potential $u^{FCH}(x)$ is the
              Hartree contribution of the initial core.
              }
      \label{fig1}
 \end{center}
\end{figure}

A schematic of the 1D model potential $U(x)$ of the nano-device is given in
Fig. \ref{fig1}. The potential is determined by $U(x) = U_0$ for $|x|\le l$
and $U(x)=V_{nd}(x)+V_{c}(x)$ otherwise. Atomic units are used
throughout the paper, where $1$ bohr $= 0.0529$ nm, $1$ hartree $= 27.2$ eV,
and $i_0=6.6 \times 10^{-3}$ A is
the atomic unit of current. The parameter $U_0$ represents the depth of
the nano-device and its length is $2l$.
We assume that the device is neutral, i.e., it contains the same number of
ions and electrons, $f_c=4$.
Therefore, for $|x| > l$, $U(x)$ is composed of the ion potential
$V_{nd}(x)=-\sum_{i=1}^{f_c} 1 /
\sqrt{a_n^2+(x-d_i)^2}$, and of the effective conductor potential,
$V_{c}(x) =U_f [2-\tanh((d + x)/ w)-\tanh((d-x)/ w)]/2$.
Here $d_i$ is the position of the $i$th nucleus, $a_n$ is the shielding
parameter of the 1D ion potential, and $U_f$ is the effective potential
in the conductor. Parameters $d$ and $w$ are used to
model the potential in the transition region between nano-device
and metal \cite{Lang92prb}.

Further the Coulomb interaction between the $f=f_c+1$ electrons is modelled by $(1/2) \sum_{i,j}^{f} V_{ee}(x_i,x_j)$
with $V_{ee}(x_i,x_j)=1/\sqrt{a^2+(x_i-x_j)^2}$. The shielding parameter $a$ removes the Coulomb singularity. As
the averaging over the dimensions transversal to the direction of electron transport results in an effective,
shielded Coulomb interaction, $a$ can be viewed to represent the transversal properties of the nano-device
\cite{Tamb99}.

For all the calculations presented here the number of electrons is fixed to $f=5$, four core electrons in the
nano-device plus one electron tunneling through the device. The wavefunction of the transport electron is given by
a Gaussian initial wavepacket, $\varphi_t$, with $1/e$-width $\alpha$, momentum $p_0$, and energy $E_0 = p_0^2/2$,
see Fig.\ \ref{fig1}). The initial distance from the tunneling device, $x_0 = -200$ bohr, is chosen large enough
so that the total multi-electron initial wavefunction can be written as an anti-symmetrized tensor product of the
electronic ground state of the nano-device, $\Psi_c$, and of the transport electron, i.e.,
$ \Psi = {\cal A} [\Psi_c(q_1,q_2,q_3,q_4) \otimes \varphi_t(q_5)]$. Here, ${\cal A}$ denotes the
anti-symmetrization operator, and $q_i = (x_i,s_i)$ ($i=1,...,5$) the space-spin coordinate.

%%%%%%%%%%%%%%%%%%%%% 2cw2eldsb.eps ****************
%%%%%%%%%%%%%%%%%%%%%%%%%%%%%%%%%%%%%%%%%%

The MCTDHF current is a combination of various single and multi-electron processes. To untangle the different
contributions and their importance we calculate the electron transport by different methods. In the MCTDHF
approach the exact wavefunction is approximated by the ansatz
\begin{equation}
\Psi = \frac{1}{\sqrt{f!}}\sum_{j_1,...,j_f}^{n}A_{j_1...j_f}(t) \varphi_{j_1}(q_{1},t)...
\varphi_{j_f}(q_{f},t) .
\label{e1}
\end{equation}
Note that the number of wavefunctions $n$ in the ansatz is larger than the number of electrons, resulting in
$\binom{n}{f}$ Slater determinants. The ansatz is inserted into the Schr\"odinger equation and the Dirac-Frenkel
variational method is used to derive a coupled system of time-dependent, nonlinear, partial differential
equations for the coefficients $A$ and wavefunctions $\varphi$.  For more information on the MCTDHF method, see
Ref.\ \onlinecite{Zanghellini04}. In the calculations the simulation interval is chosen from $x=-400$ bohr to
$x=400$ bohr with the number of grid points $N = 800$. Convergence is obtained for $n=11$ wavefunctions.
Increasing $n$ and $N$ changes the electron density by less than 0.1\%.

The second method used in our investigation is time-dependent Hartree Fock (TDHF), which is MCTDHF for $f=n$. The
third and fourth methods are frozen core methods, where the correlated ground state wavefunction of the nano-device
is calculated by MCTDHF and assumed to remain frozen during the transport of the conducting electron. In dependence
on whether exchange effects between ground state and conducting electron are taken into account, these approaches
are termed frozen core Hartree (FCH) and frozen core Hartree Fock (FCHF). The frozen core approximation reduces the
multi-electron Schr\"odinger equation to a single-electron equation given by
\begin{eqnarray}
&&i\partial_t\phi(x,t)=h\phi(x,t)+\int V_{ee}(x',x)\rho_c(x')dx'\phi(x,t)
  \nonumber \\
&&
  -\int \Gamma_c^{(1)}(q'|q) V_{ee}(x',x) \phi(x',t)\delta_{s,s'}dq'
  \nonumber \\
&&
  -\int \Gamma_c^{(1)}(q'|q) h(x') \phi(x',t)\delta_{s,s'}dq'
  \nonumber \\
&&
 -2\int \Gamma_c^{(2)}(q',q''|q',q)V_{ee}(x',x'')\phi(x'',t)\delta_{s,s''}dq'dq'',
    \nonumber\\
\label{FCHFEq}
\end{eqnarray}
where $s$ is the spin index of the conducting electron, $\rho_c(x)=\int \Gamma_c^{(1)}(q|q)ds$, and $h=p^2/2+U(x)$.
The core electron reduced density matrices are defined as $\Gamma_c^{(1)}(q'_1|q_1)=[2/(f_c-1)]\int
\Gamma_c^{(2)}(q'_1,q_2|q_1,q_2) dq_2$ and $\Gamma_c^{(2)}(q'_1,q'_2|q_1,q_2)=\big{(}^{f_c}_2\big{)} \int
\Psi_c^*(q'_1,q'_2,q_3,...,q_{f_c}) \Psi_c(q_1,q_2,q_3,...,q_{f_c})dq_3...dq_{f_c}$ \cite{Lowdin}. The first two
terms on the right hand side of Eq.\ (\ref {FCHFEq}) give the frozen core Hartree (FCH) approximation. The last
three terms account for the ground state correlation of the nano-device and the exchange effect between bound and
tunneling electrons.
In order to keep the problem as simple as possible, we focus here on
the field-free scattering of the electron from the nano-device. In the
weak field limit, where the field induced distortion of the bound state
is weak, the change of the initial energy of the incoming electron $E_0$
is equivalent to applying a bias voltage $V \approx E_0$
determined by the chemical potential of the two leads.
One atomic units of energy change corresponds to a bias
change of $27.2V$.

\begin{figure}
  \begin{center}
      \includegraphics[width=8.5cm,height=6.0cm]{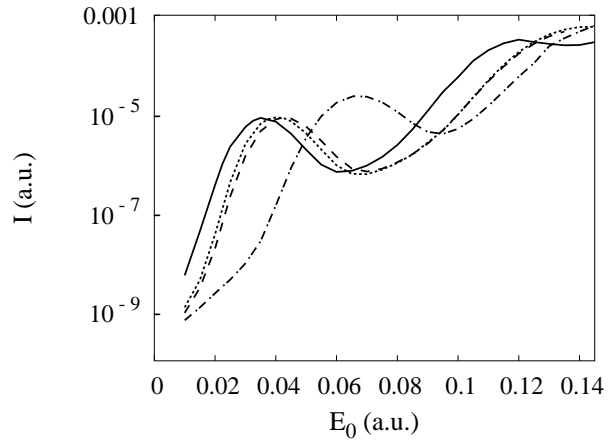}%{figs45ie0.eps}
      \caption{
      Average current versus energy $E_0$,
      obtained by  MCTDHF (full), TDHF (dotted), FCHF (dashed), and
      FCH (dash-dotted). Parameters are: $f_c=4$, $a=4.5$ bohr,
      $2l=10$ bohr,  $d=11.0$ bohr, $w=0.8$ bohr, $\alpha=40$ bohr,
      $x_0=-200$ bohr,
      $U_0=-0.92$ hartree, %$a_n=2.76$, $d_n=3.3$,
      $U_f=-0.18$ hartree ($-5$ eV),
      and $I_p=0.382$ hartree ($10.4$ eV).
      %For details see text.
      }
      \label{fig2}
 \end{center}
\end{figure}

\begin{figure}
  \begin{center}
      \includegraphics[width=8.5cm,height=6.0cm]{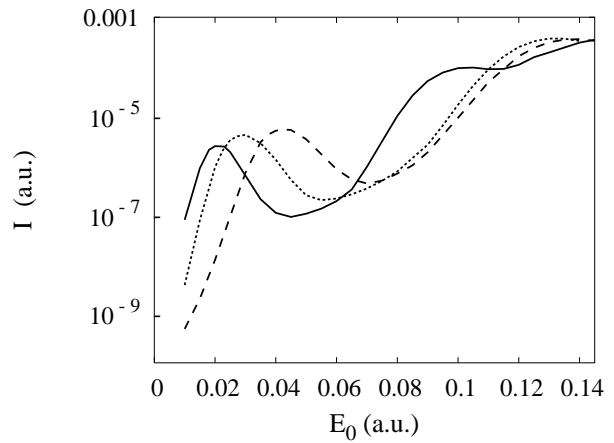}%{figs3ie0.eps}
      \caption{
      Average current versus energy $E_0$,
      obtained by  MCTDHF (full), TDHF (dotted), and FCHF (dashed).
      Parameters are the same as those given in Fig.\ \ref{fig2},
      except for $a=3.0$ bohr. The following parameters had also to
      be modified to keep the ionization potential
      $I_p=0.388$ hartree  ($10.6$ eV) and resonance
      structure as close as possible to the system in Fig. 2:
      $2l=12$ bohr, and $U_0=-0.97$ hartree.}
      \label{fig3}
 \end{center}
\end{figure}

In Figs.\ \ref{fig2} and\ \ref{fig3} we have plotted the average current
$I$ flowing through the nano-device as a function of the energy of the
incoming electron. The parameters have been chosen to reflect
experimental systems, see the figure caption. The device size of around
$0.5$ nm, the ionization potential of $10$ eV, and the
Fermi energy of $-5$eV are typical values found in experimental
nano-molecule systems, such as in the
Au/dithiolated-benzene/Au junction \cite{Reed}.
The average current $I(x) =<\tilde{i}(x,t)>$ is
calculated between the initial time and the time when the reflected
pulse returns to its initial position. The time
dependent tunneling current $\tilde{i}(x,t)$ is obtained by
\begin{eqnarray}
&& \tilde{i}(x,t) =\frac{f}{2i}\nonumber \\
&&\times \sum_{j,k}^{n}
  \tilde{\rho}_{j,k}[\varphi^*_j(x,t)\partial_x \varphi_k(x,t)
  -\varphi^*_k(x,t)\partial_x \varphi_j(x,t)],\ \  \
\label{qflux}
\end{eqnarray}
where
$\tilde{\rho}_{j,k} =\sum_{i_2,...i_f}^{n}A^*_{j,i_2...i_f}A_{k,i_2...i_f}$ denotes the density matrix and $I(x)$
is calculated at $x=80$ bohr. The electron energy $E_0$ is measured with reference to $U_f$.

The major difference between Figs.\ \ref{fig2} and\ \ref{fig3} is the use
of different shielding parameters $a=4.5$ bohr and $a=3.0$ bohr,
respectively. The decreasing value of $a$ reflects an increasing
electron-electron interaction strength. Although it is difficult to
assign the 1D Coulomb correlation strength resulting from different
values of $a$ to specific devices, some qualitative considerations can
be made. We believe that $a \approx 2$ to $5$
represents best the 3D experimental situation. On the one hand, calculations
for $a=7$ show hardly any interaction between the transport and the core
electrons. On the other hand, for $a < 2$, the electron repulsion becomes
too strong. As a result, the ionization potential increases and can no
longer be fitted to values typical of experimental devices.

The MCTDHF, TDHF, and FCHF results are depicted by the full, dotted, and dashed lines, respectively. The dash-dotted
line in Fig.\ \ref{fig2} represents the FCH result. All curves show the same general trend. Whereas the structure of
the curves is determined by the resonances, their growth with increasing $E_0$ depends mainly on the height and
thickness of the tunneling barrier. There are two resonances in the ranges $E_0=0.02 \sim 0.04$ and $E_0=0.1 \sim
0.14$. The strength and width of the second resonance is more pronounced mainly due to the weaker tunneling barrier
for higher $E_0$.

The comparison of the various approaches reveals the role of exchange and correlation effects in determining the
current through the model nano-system. The FCH approach neglects two important effects: (i) exchange interaction
between conducting and bound electrons; (ii) interaction between tunneling and core electrons in the sense that
the tunneling electron induces a perturbation to the bound electrons of the nano-system. The resulting potential
change modifies the tunneling current. In order to capture this effect, the frozen core approximation has to be
removed. In the FCHF approximation the exchange interaction is taken into account, whereas approximation (ii)
remains. In the TDHF approach part of the limitation (ii) is removed by making the wavefunction, that governs the
bound electron dynamics, time-dependent. However, TDHF does still not fully account for the cross-talk between
conduction and bound electrons, as correlation is neglected by the product ansatz of single-electron wavefunctions.
Finally, in MCTDHF all effects are treated properly.

%%%%%%%%%%%%%%%%% figs45ie0 %%%%%%%%%%%%%%%%%%%%%%%%
%%%%%%%%%%%%%%%%%%%%%%%%%%%%%%%%%%%%%%%%%%%%%%%%
%/LT1 { PL [4 dl 2 dl] 0 1 0 DL } def!dot: /LT1 { PL [10 dl 10 dl] 0 1 0 DL } def
%/LT2 { PL [2 dl 3 dl] 0 0 1 DL } def!dash:/LT2 { PL [3 dl 3 dl] 0 0 1 DL } def
%              /LT5 { PL [4 dl 3 dl 1 dl 3 dl] 1 1 0 DL } def
%dash-dotted: %/LT5 { PL [10 dl 6 dl 2 dl 3 dl] 1 1 0 DL } def
%%%%%%%%%%%%%%%%% figs3ie0 %%%%%%%%%%%%%%%%%%%%%%%%
%%%%%%%%%%%%%%%%%%%%%%%%%%%%%%%%%%%%%%%%%%%%%%%%

Figure\ \ref{fig2} shows that FCH introduces a significant error with regard to the position of the resonances,
as can be seen from the first resonance. As a result of the large width of the second resonance, the difference
between FCH and MCTDHF is blurred. In Fig.\ \ref{fig3}, FCH was not plotted as the error in resonance position
becomes of the order of the difference between first and second resonances. Comparison of FCH and FCHF in
Fig.\ \ref{fig2} shows that a substantial part of the resonance shift comes from the neglect of the exchange
effect.

The remaining difference can be attributed to the perturbation of the electrons in the nano-device caused by the
tunneling electron. The resulting change of the electronic core modifies the position of the resonances and
therewith the tunneling current. A comparison of the FCHF, TDHF and MCTDHF in Fig.\ \ref{fig3} demonstrates the
importance of correlation in the interaction between bound and tunneling electrons \cite{sai05}. Although the
TDHF approach performs better than FCHF, it is in most places off by an order of magnitude from the MCTDHF
calculation. This highlights the fact that during tunneling a highly correlated state between core and transport
electrons builds up. The amount of correlation strongly depends on the electron-electron interaction strength,
as can be seen from a comparison between Figs.\ \ref{fig2} and\ \ref{fig3}.

Finally,  over most of the energy range ($E_0 \le 0.12$) the electron transport is elastic. The nano-device returns
to its ground state after the conducting electron has passed. Only for energies ($E_0 > 0.12$) we find evidence of
excited state population. The excitation causes a suppression of the resonance at $E_0 = 0.12$ in Fig.\ \ref{fig2}.
This "strong-field" limit is not analyzed further here.

%%%%%%%%%%%%%%%%% 4cs345psum.eps  %%%%%%%%%%%%%%%%%%%%%%%%
\begin{figure}
  \begin{center}
      \includegraphics[width=8.5cm,height=6.0cm]{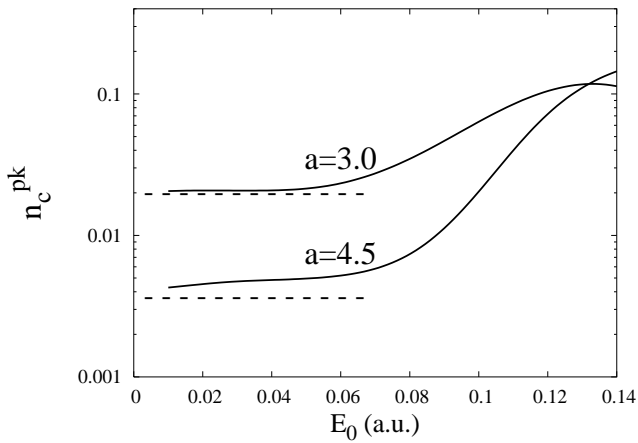}%{4cs345psum.eps}
      \caption{
      Dashed lines: correlations of the initial bound states,
      $n_c(a,t=0)$.
      Solid curves:
      overall correlations measured at the tunneling peak of the
      nano-devices, $n_c^{pk}(a)$, % with $a=3.0$ and $a=4.5$.
      the peak value of $n_c(a, t)$. The difference between solid and
      dashed lines is the contibution due to the interaction between tunneling and
      core electrons.
      }
      \label{fig4}
 \end{center}
\end{figure}
%%%%%%%%%%%%%%%%%%%%%%%%%%%%%%%%%%%%%%%%%%%%%%%%

One way to characterize the build up of Coulomb correlation between tunneling and bound electrons is to
consider the occupation number of the natural orbitals, which are the eigenvalues of the density matrix
$\tilde{\rho}$ \cite{Lowdin}. The number of natural orbitals is $n > f$ and they are ordered with respect
to their eigenvalue. The lowest orbital refers here to the largest occupation number. In the uncorrelated
HF limit only the lowest $f$ orbitals are occupied. Therefore the Coulomb correlation can be measured by
%\begin{eqnarray}
$n_{c}(a,t)=\sum_{i=f+1}^{n} \lambda_i(a,t)/f$,
%\label{nc}
%\end{eqnarray}
which is the occupation number of the natural orbitals with $i> f$. Here, $\sum_{i=1}^{n}\lambda_i(a,t)=f$
$\lambda_i(a,t)> \lambda_{i+1}(a,t)$, and $0 \leq n_c(a,t) \leq 1-f/n$.

The correlation measure is plotted in Fig.\ \ref{fig4} for the parameters of Figs. \ref{fig2} and \ref{fig3}.
Up to $E_0 \approx 0.06$ the correlation introduced by the tunneling electron is smaller than the initial core
correlation $n_c(a,t=0)$, which is independent of the value of $E_0$. In this range the physics is dominated
by the bound electrons that determine the position of the resonances. For larger energies, the interaction
between bound electrons and tunneling electron adds a significant amount of correlation to the system. The
sharp increase of $n_c(a,t)$ agrees with the large difference between MCTDHF and the other approaches in Figs.\
\ref{fig2} and \ref{fig3}.

%{0}%{abc}
\end{document}